\documentclass[preprint]{elsarticle}

\usepackage[top=2cm, bottom=2cm, left=3cm, right=2cm]{geometry}

\begin{document}

\begin{frontmatter}

\title{Large-scale distribution of cosmic rays in right ascension\\ as observed by the Yakutsk array at energies above $10^{18}$ eV}

\author{A.A. Ivanov}
\ead{ivanov@ikfia.ysn.ru}
\author{A.D. Krasilnikov}
\author{M.I. Pravdin}
\author{A.V. Sabourov}

\address{Shafer Institute for Cosmophysical Research and Aeronomy,
31 Lenin Avenue, Yakutsk 677980, Russia}

\begin{abstract}
We present the results of searches for anisotropy in the right ascension ($RA$) distribution of arrival directions of cosmic rays (CRs) detected with the Yakutsk array during the 1974--2008 observational period in the energy range above $10^{18}$ eV. Two methods of analysis are applied to two sub-samples of the data. Particularly, estimations of the first and second harmonic amplitudes are given, as well as the first harmonic phase in adjacent energy intervals. Analysis of variance demonstrates a significant contraction of the minimal width of the $RA$ distribution in the energy bin $(10^{19},1.78\times10^{19})$ eV with respect to the isotropic distribution, which may be attributed to a possible source of CRs within the interval $RA\in(15^0,45^0)$.
\end{abstract}

\end{frontmatter}


\section{Introduction}
A conventional approach to shed light on the origin of cosmic rays (CRs) is to search for anisotropy in the arrival directions distribution. A number of attempts have been made to find excess fluxes of CRs correlated with large-scale structures in the nearby Universe, resulting in several indications of possible anisotropic effects, but none of these effects have been confirmed independently (some examples can be found in Refs.\ \cite{Hillas,Sommers,Malfa}).

By restricting the data under analysis exclusively to ground-based arrays, an essentially uniform exposure in the right ascension can be obtained. Under this restriction, in the present paper we examine the data obtained from scintillation counters--surface detectors of the Yakutsk array. The results of previous analysis of the $RA$ distribution of extensive air shower (EAS) primaries detected within the time period from January 1974 to May 2000 were published in Refs.\ \cite{Hamburg,Kashiwa,Wavelet}. We have now analyzed the extended dataset up to June 2008. Additional data include, specifically, 7598 EAS events above the threshold energy 1 EeV (=$10^{18}$ eV) \cite{Rio}.

The main aim of this analysis was to extend the time series of the data by using an observational time that was as long as possible. These efforts were aimed at testing the previous indications of possible anisotropy in the arrival directions of CRs in the extended dataset using different methods.

\begin{figure*}[t]\centering
\includegraphics[width=0.4\textwidth,angle=90]{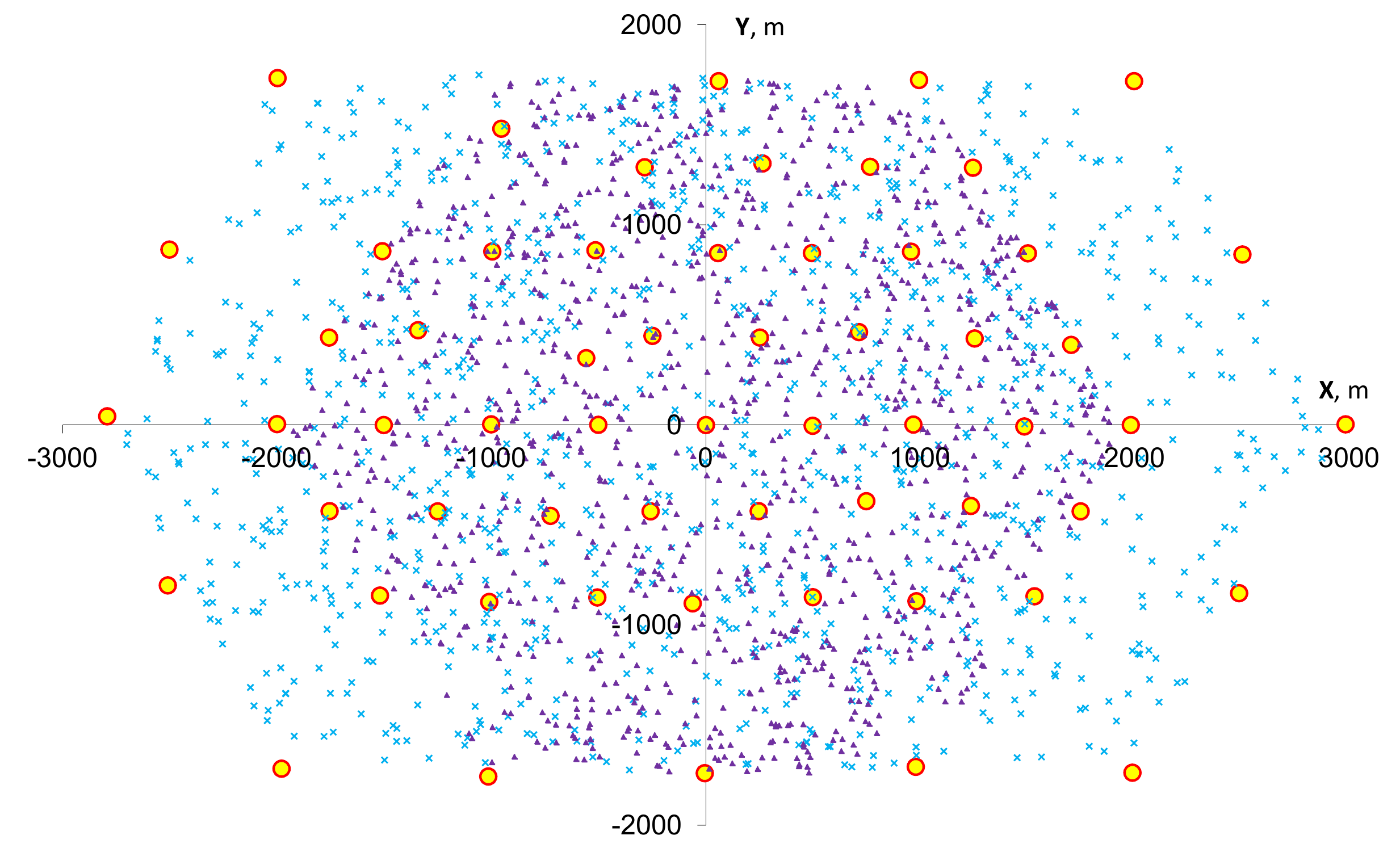}
  \caption{Arrangement of the Yakutsk array detectors (stations forming trigger-500 and trigger-1000; circles) and the shower cores selected before (crosses) and after 1990 (triangles).}
\label{Fig:Map}\end{figure*}


\section{The Yakutsk array experiment and data selection for analysis}
The Yakutsk array site is located near Oktyomtsy, the satellite village of Yakutsk, at geographical coordinates $61.7^0N,129.4^0E$ and at a mean altitude of 105 m above sea level. At present, it consists of 58 ground-based and 4 underground scintillation counters to measure charged particles (electrons and muons), and 48 detectors of the air Cherenkov light. During its 40-year lifetime, the array has been reconfigured several times. Before 1990, the total area covered by detectors was at its maximum ($S\sim17$ km$^2$); now, it is $8.2$ km$^2$.

EAS events were selected from the background using a two-level trigger of detector signals: The first level is a coincidence of signals from two scintillation counters in a station within 2 $\mu$s; the second level is a coincidence of signals from at least three nearby stations (not lined up) within 40 $\mu$s. Stations spaced $\sim500$ m and $\sim1000$ m form the so-called trigger-500 and trigger-1000, respectively. Over the entire observation period, more than $10^6$ showers of primary energy above $10^{15}$ eV were selected. Fig.\ \ref{Fig:Map} shows the positions of detectors of the Yakutsk array and the core positions of EASs with $E>3$ EeV. Only scintillation counters of the array with spacing 500 and 1000 m are shown. Other types of detectors are described, for example, in Refs.\ \cite{Kashiwa,NJP,MSU}.

\begin{figure*}[b]\centering
\includegraphics[width=0.49\textwidth]{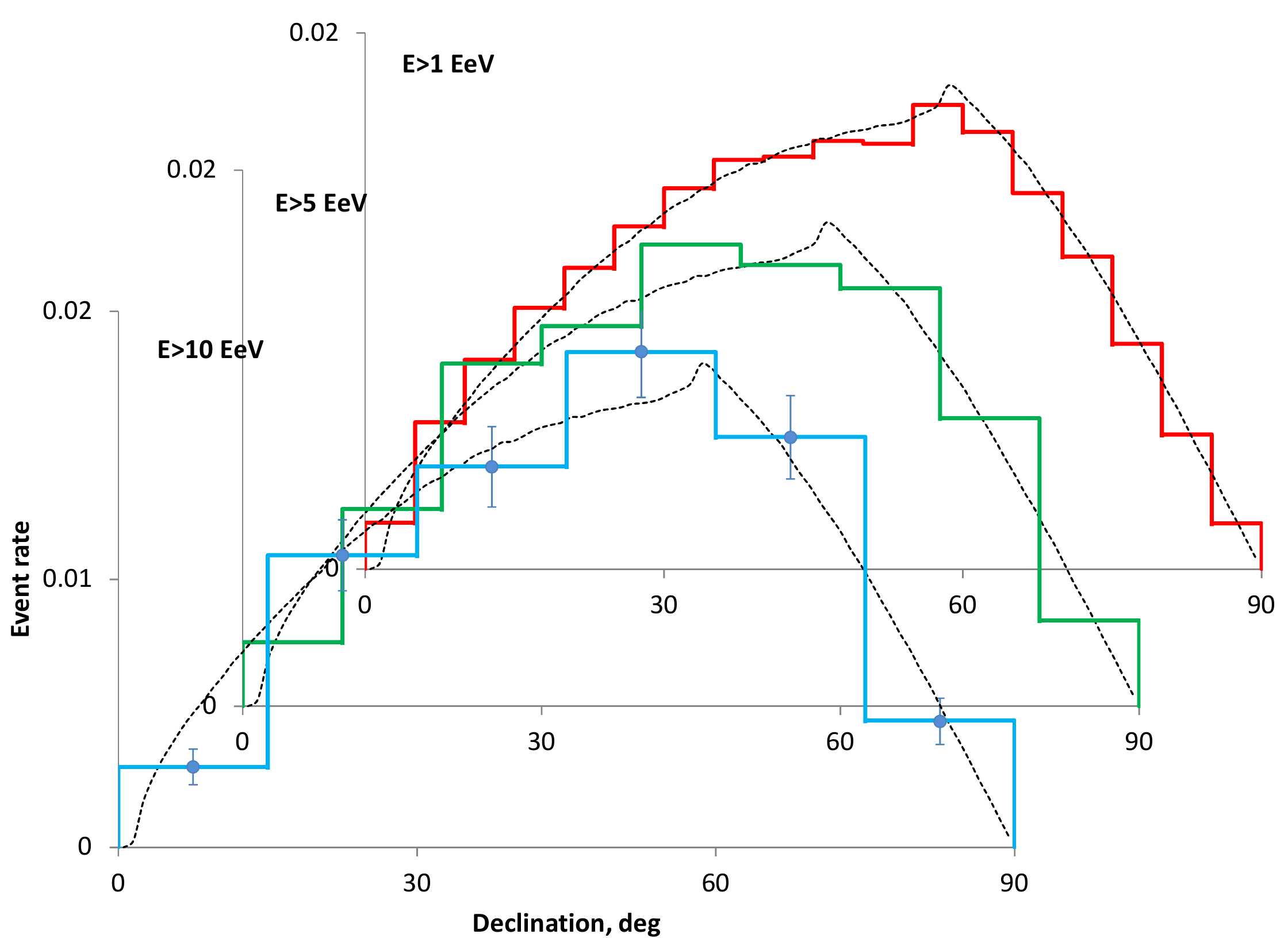}
\includegraphics[width=0.49\textwidth]{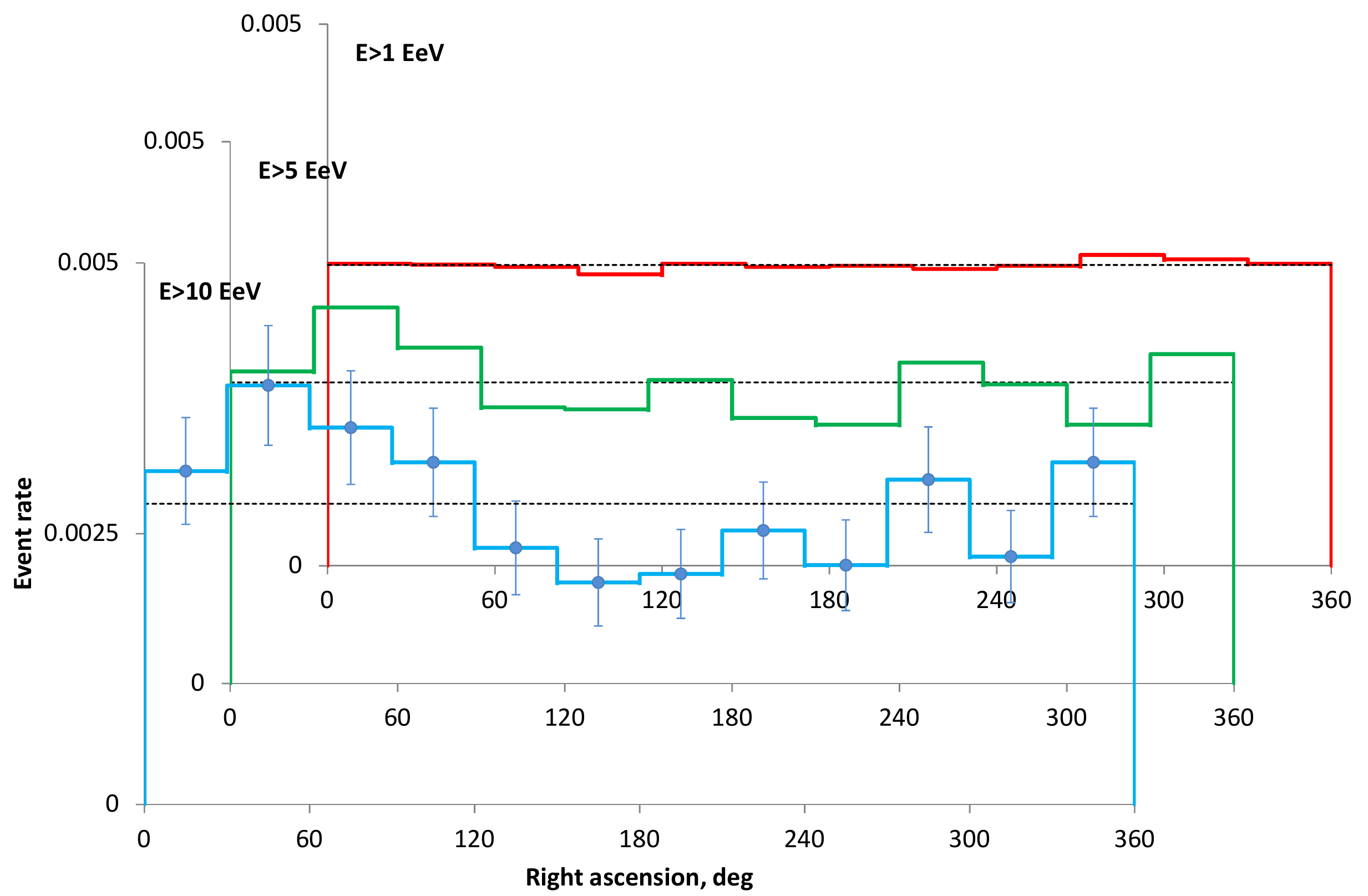}
  \caption{Arrival directions in equatorial coordinates. Extensive air showers were selected with three threshold energies (histograms). Isotropic distributions are shown as dotted curves.}
\label{Fig:EqFig}\end{figure*}

The shower core coordinates were located fitting the lateral distribution of particle densities by the Greisen-type function. Core location errors are $\sim30$ m for trigger-500 and $\sim50$ m for trigger-1000 events.

Arrival angles of the EAS primary particles were calculated in the plane shower front approximation using detection times at the stations. A clock pulse transmitter at the center of the array provided pulse timing to 100 ns accuracy. Errors in arrival angles depend on the primary energy decreasing from $\sim7^0$ at $E=1$ EeV to $\sim3^0$ above $E=10$ EeV. More detailed information can be found in Refs.\ \cite{NJP,MSU,Site}.

In this work, a sample of the analyzed dataset consisted of EAS events detected in the period January 1974 -- June 2008 within the array area, with energy above $1$ EeV, at zenith angles $\theta<60^0$.

The energy estimation method was based on the total flux measurement of the air Cherenkov light and the number of electrons and muons at observation level~\cite{NJP,JETP,IzvRAN}. To unify the energies of showers detected in a variety of years, we used the same $S_{600}$-to-energy relationship and attenuation length as in Ref.\ \cite{Kashiwa}. The number of EAS events selected was 43710 at $E>1$ EeV. The energy estimation error was approximately $35\%$ for the showers with axes within the array area \cite{IzvRAN}. The systematic error in the energy estimation procedure of the Yakutsk array, as in the case of other giant arrays, could be corrected by a specific factor \cite{APJ,CERN}.

Fig.\ \ref{Fig:EqFig} shows the distribution of arrival directions of selected showers in the equatorial system compared with expected isotropic distributions.


\section{Harmonic analysis}

We used the Rayleigh formalism in harmonic analysis of the $RA$ distribution of CR arrival directions \cite{Rayleigh}. If the large-scale anisotropy is present in the arrival directions of CRs, it can be expressed using the anisotropy coefficient for the intensity variation:
$$
A=\frac{I_{max}-I_{min}}{I_{max}+I_{min}},
$$
where $I_{max}$ and $I_{min}$ are the maximum and minimum of CR intensity, respectively, as a function of the arrival angle, $\alpha$. In the simplest cases with one source at $\alpha_0$ (for example, $I_{max}$ from the Galactic center, $I_{min}$ from the anticenter), the distribution can be described by a cosine wave $I(\alpha)=1+A\cos(\alpha-\alpha_0)$, or with the two opposing sources as $I(\alpha)=1+A\cos(2\alpha-2\alpha_0)$ (for example, $I_{max}$ from the Galactic arm ``in'' and ``out'' directions, while $I_{min}$ is from perpendicular directions). In the first case, the coefficient is denoted $A_1$, which is the amplitude of the first harmonic, and in the second case, the coefficient is denoted $A_2$, which is the amplitude of the second harmonic. In the general case of harmonic expansion of an arbitrary distribution:
\begin{equation}
I(\alpha)=1+\sum_{k=1}^\infty A_k\cos(k\alpha-k\alpha_0).
\end{equation}
If the observed distribution of arrival angles, $\alpha_i$, is given by a sum of delta functions $f(\alpha)=\sum_{i=1}^N \delta(\alpha-\alpha_i)$, then
\begin{equation}
A_k=\sqrt{a_k^2+b_k^2},\hspace{3mm} \alpha_0=\arctan(b_k/a_k),
\end{equation}
where $a_k=\frac{2}{N}\sum_i\cos(k\alpha_i),\hspace{3mm}b_k=\frac{2}{N}\sum_i\sin(k\alpha_i)$.

\begin{figure*}[t]\centering
\includegraphics[width=0.45\textwidth]{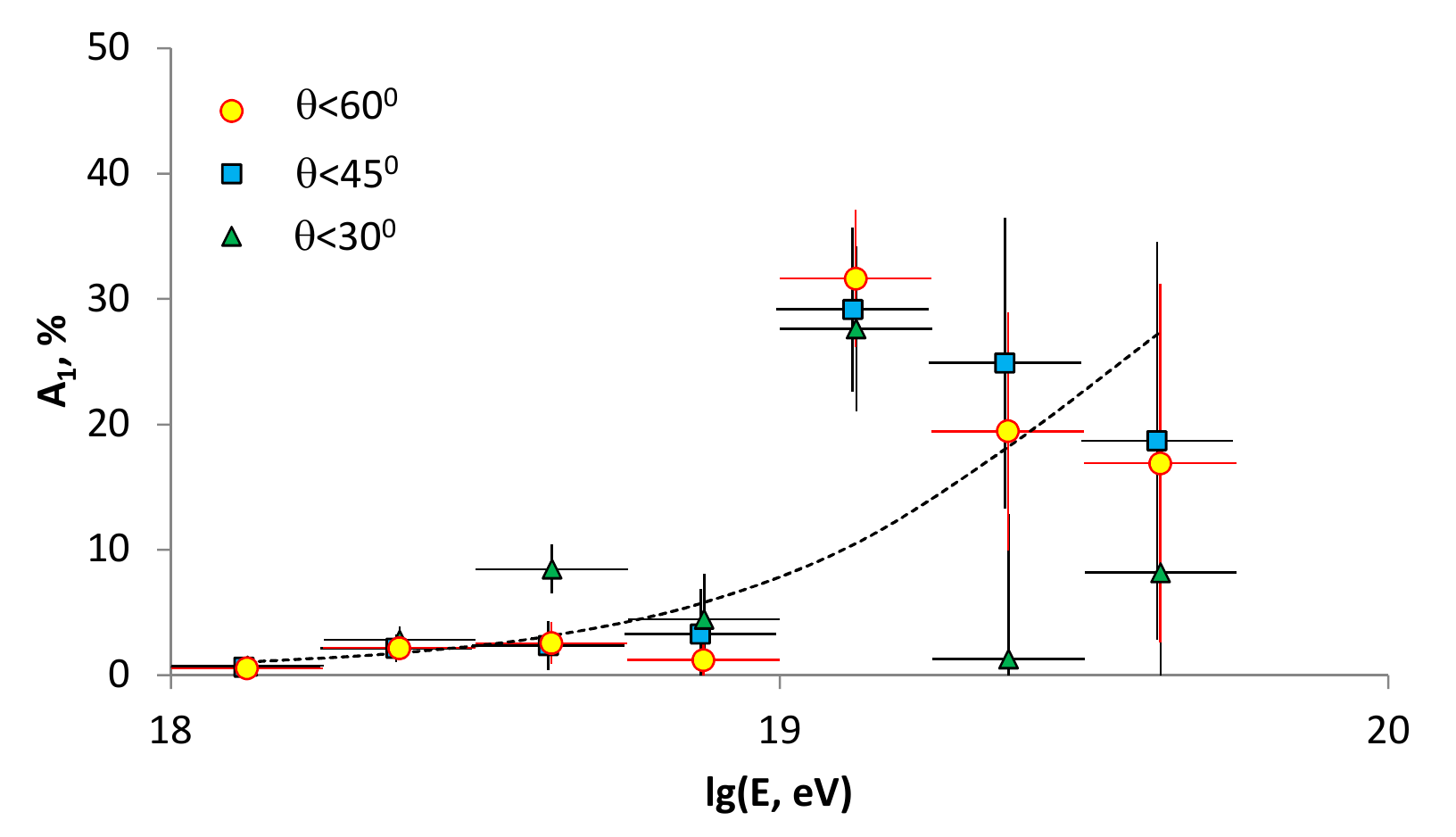}
\includegraphics[width=0.45\textwidth]{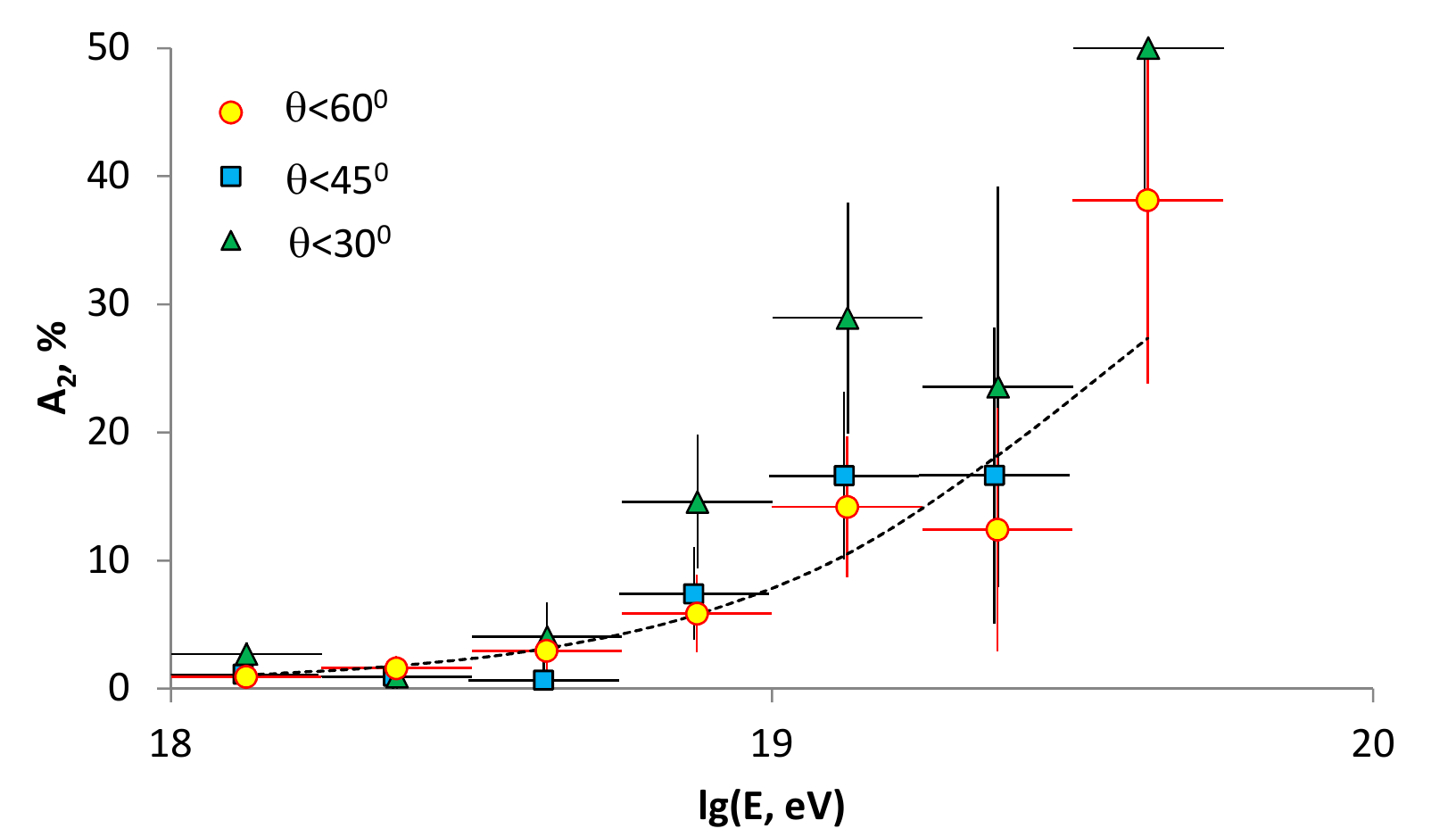}
  \caption{The first (left panel) and second (right panel) harmonic amplitudes in the right ascension distribution of cosmic rays detected in the period 1974--2008 at energies above $1$ EeV. Amplitudes of the expected isotropic distribution ($\theta<60^0$) are indicated as dashed curves. Vertical bars are statistical errors; horizontal bars indicate energy bins.}
\label{Fig:Harmonics}\end{figure*}

Average values of the amplitudes are not zero even in the case of isotropic distribution, and converge to zero with $N\longrightarrow\infty$ (see Appendix A):
\begin{equation}
\overline{A_k^0}=\sqrt{\frac{\pi}{N}},\hspace{3mm} \sigma(A_k^0)=\sqrt{\frac{4-\pi}{N}}.
\end{equation}
The probability that isotropic amplitudes will be larger than the given $A_k$ is
\begin{equation}
P(>A_k)=\exp(-\frac{NA_k^2}{4}).
\end{equation}

Earth's rotation enables celestial regions to be represented as $RA$ distributions. Scintillation counters of the Yakutsk array have a 24-hour duty cycle, resulting in almost uniform directional exposure. Small deviations from the uniformity are caused by maintenance of the detectors (mainly during working days) and summertime shutdowns of the array due to thunderstorms.

Diurnal and seasonal variations of the array exposure and atmospheric conditions result in spurious amplitude $(0.45\pm0.55)\%$ \cite{Pravdin}. With datasets of the size described in section 2, we were able to determine only those amplitudes that were well above $A_k^0\simeq0.85\%$ (Eq. 3). Therefore, in the energy region $E>1$ EeV, we neglected the effect of diurnal and seasonal variations of the array exposure in the analyzed dataset.

\begin{table}[t]
\begin{center}
\caption{Yakutsk array observed amplitudes in energy bins, $A_k$, and the probability that expected isotropic amplitudes will exceed observed amplitudes, $P(>A_k)$. $\theta\in(0^0,60^0)$.}
\begin{tabular}{|l|r|r|r|r|}
\hline Energy bins, & $A_1$, & $P(>A_1)$, & $A_2$, & $P(>A_2)$,\\
       $\lg(E, eV)$ &   \%   &   \%       &   \%   & \% \\ \hline
18.0-18.5 &  0.95  & 41.42 &  0.71  & 60.10 \\ \hline
18.5-19.0 &  2.08  & 64.03 &  0.99  & 90.34 \\ \hline
19.0-19.5 & 28.54  &  0.04 & 11.78  & 26.75 \\ \hline
19.5-20.0 & 16.89  & 74.10 & 38.13  & 21.73 \\ \hline
\end{tabular}
\end{center}
\end{table}

The efficiency of array detection is also affected by the geomagnetic field. Because the trajectories of charged CRs are curved in a magnetic field, the distribution of particles becomes oval along the Lorentz force. Owing to the steepness of the energy spectrum, such an azimuthal dependence translates into azimuthal modulation of the EAS event rate for a given zenith angle and particle density \cite{Geo1}. Observed distributions of arrival directions in a horizontal system may also be distorted because of the geomagnetic effect. Fortunately, the $RA$ distribution is not affected, because of diurnal smearing by the Earth's rotation \cite{Geo2}. As a result, we ignored the geomagnetic field effect at $E>1$ EeV.

Fig.\ \ref{Fig:Harmonics} shows the resultant harmonic amplitudes of the data in three zenith angle intervals. We did not derive higher harmonics, $A_k, k>2$, because we considered these to be insufficiently large at the angular scale. Observed amplitudes increased with energy, for both the first and the second harmonics, but the effect was statistically insignificant against the background of ``isotropic'' amplitudes  increasing because of the number of events decreasing with energy. Table 1 shows the probability that the amplitude of isotropic arrival directions will be larger than the observed amplitude by chance.

A $3.3\sigma$ excess of the first harmonic amplitude over isotropic expectation was found in the energy bin $(10,17.8)$ EeV. Variation of zenith angle threshold did not eliminate the effect (Fig.\ \ref{Fig:Harmonics}) as well as doubling of the bin width (Table 1). It is interesting to note that the amplitude and phase of the first harmonic in this energy interval measured before 2000 and throughout the period had practically the same value \cite{Hamburg,Wavelet}.

\begin{figure}[b]\centering
\includegraphics[width=0.45\textwidth]{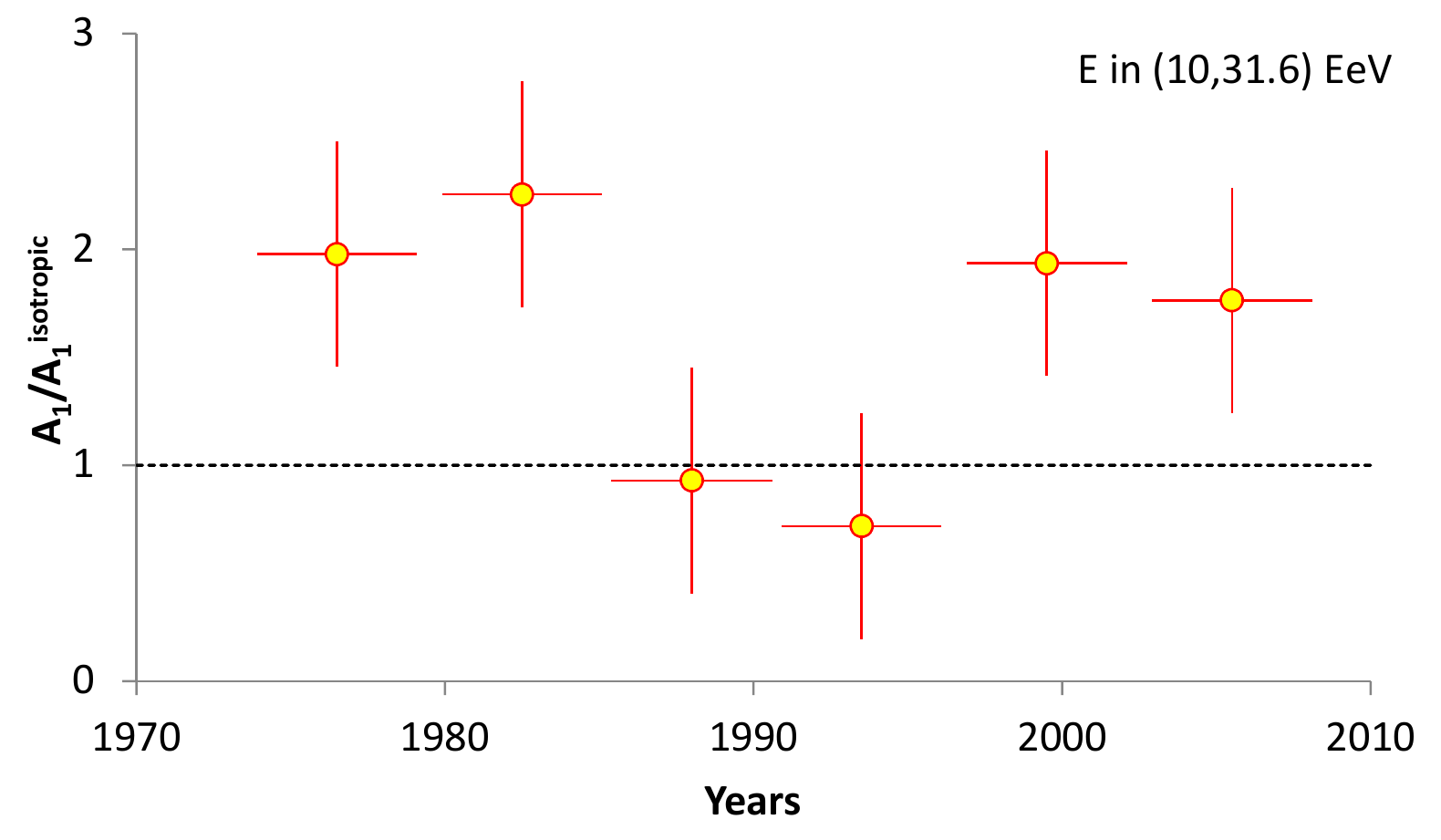}
\includegraphics[width=0.45\textwidth]{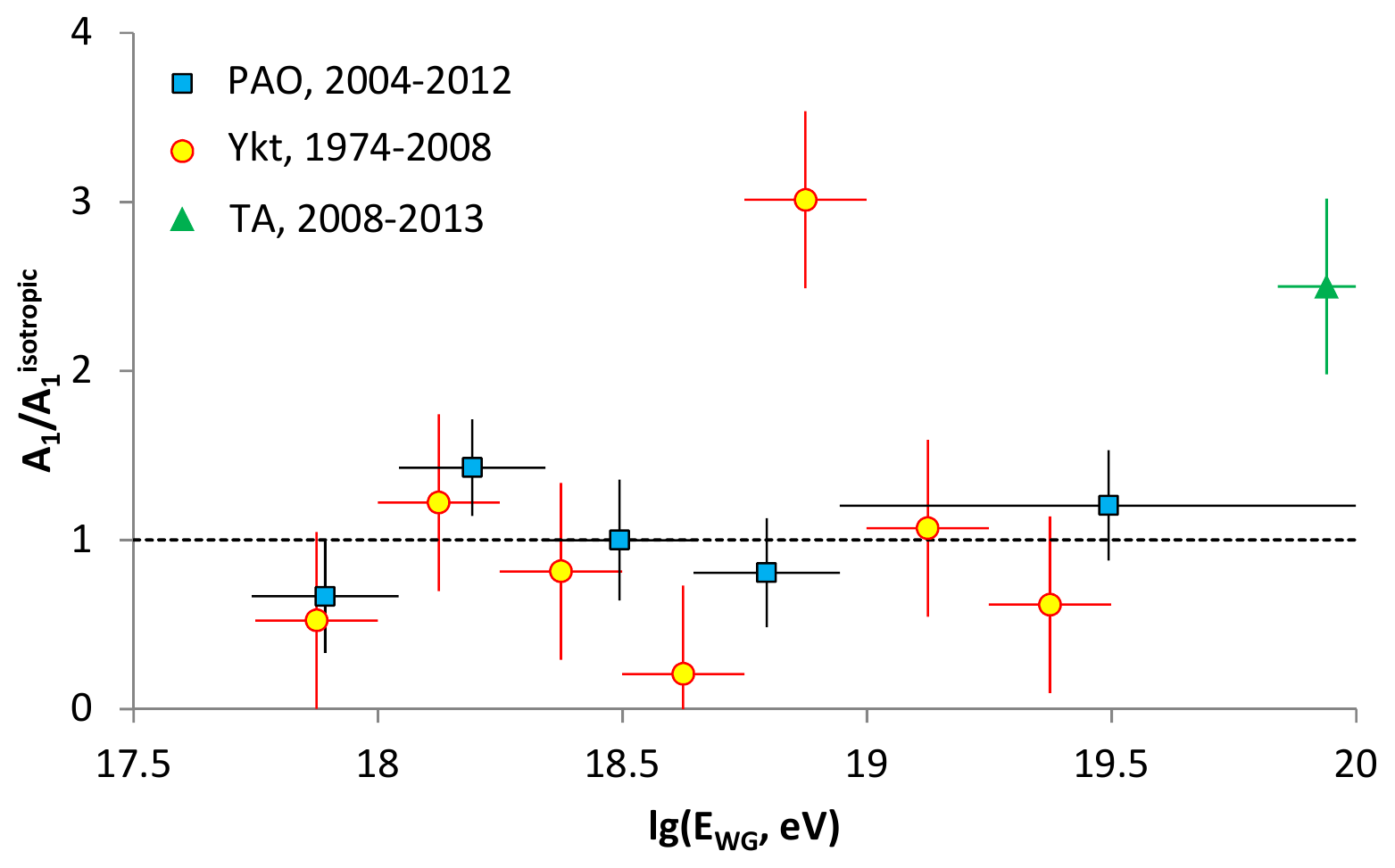}
  \caption{Ratio of observed and expected amplitudes of the first harmonic. Left panel: in six-year intervals of observation at Yakutsk. Right panel: comparison of the Yakutsk array (Ykt), Pierre Auger Observatory (PAO), and Telescope Array (TA) results. Energy, $E_{WG}$, is scaled using Working Group factors: 0.56 for Yakutsk, 0.91 for TA, and 1.1 for PAO energy estimations \cite{CERN}.}
\label{Fig:Years}\end{figure}

Time variation of $A_1(10<E<31.6)$ observed by the Yakutsk array in six-year intervals is shown in Fig.\ \ref{Fig:Years}, left panel. Anisotropic amplitudes were observed before 1985 and after 1995. The variation of amplitude can be attributed to the fluctuations of the EAS event number in reduced intervals of years.

The statistical significance of the excess flux found {\it a posteriori} in a particular energy bin should be estimated using the penalty factor that is at least equal to the number of independent bins. In this case, we consider an excess to be equally probable by chance in any of the bins.

An alternative approach is to divide the observational data into two sub-samples: EAS events detected before and after 2000. In this case, we are basing on the analysis of the Yakutsk array data consisting of two independent parts -- first results (1974-2000) are published in 2001 \cite{Hamburg}, and the second part (2000-2008) is published in 2013 \cite{Rio}. The first group was used to find the energy interval where the amplitude exceeds isotropic expectation, i.e., $E\in(10,31.6)$ EeV, where the first harmonic amplitude is $(26.4\pm 8)\%$ with a chance probability $0.004$. The second group was used to calculate the significance of the excess without the need for any statistical penalties. Data after 2000 \cite{Rio}, however, resulted in the amplitude $A_1=(41.4\pm 13)\%$ with a chance probability $P(>A_1)=0.118$ in the same energy interval, indicating no possibility of anisotropy as suggested by the first harmonic amplitude in the $RA$ distribution.

No statistically significant deviations of harmonic amplitudes were found in the data from the Pierre Auger Observatory (PAO) \cite{PAOphase} and Telescope Array (TA) \cite{TA}. The data of PAO and TA are compared with our results in Fig.\ \ref{Fig:Years}, right panel. The TA collaboration found no deviation from isotropy with energy thresholds 10 EeV and 40 EeV. However, in the highest energy bin $E>57$ EeV, they claim observation of a hotspot, with a statistical significance of 5.1$\sigma$, centered at $RA=146.7^0$, $Dec=43.2^0$ \cite{TA}.

Another item of interest can be found in the considerations of the PAO collaboration \cite{PAOphase}. It was noted that their phase measurements in adjacent energy intervals did not appear to be randomly distributed, but rather indicated a smooth transition between a common phase $270^0$ consistent with a Galactic center region below $1$ EeV and another phase consistent with the $RA$ of the anticenter above $5$ EeV. This is potentially interesting, because with a real underlying anisotropy, a consistency of the phase measurements in ordered energy intervals is indeed expected to be revealed with a smaller number of events than required to detect the amplitude with high statistical significance \cite{PAOphase}.

\begin{figure}[t]\centering
\includegraphics[width=0.45\textwidth]{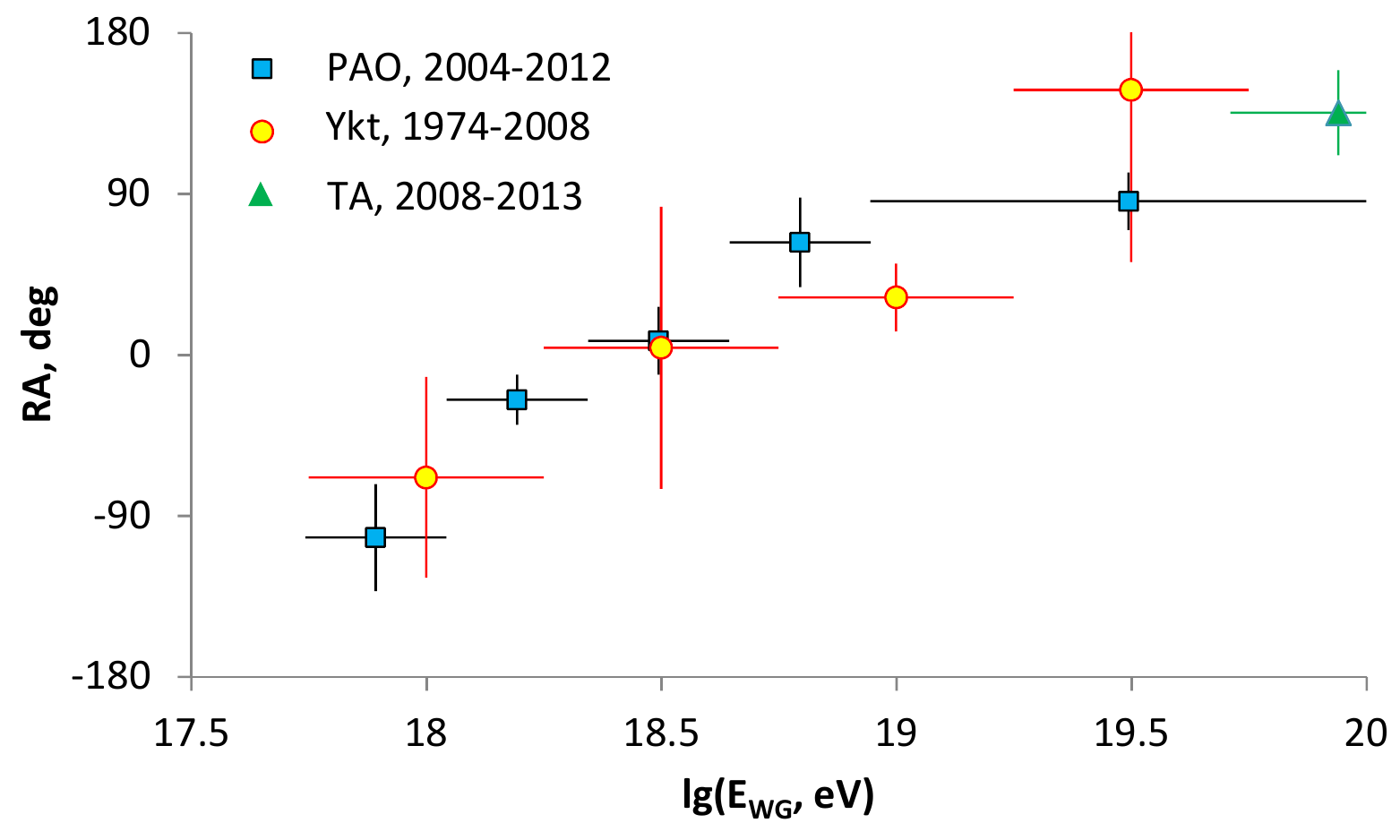}
\includegraphics[width=0.45\textwidth]{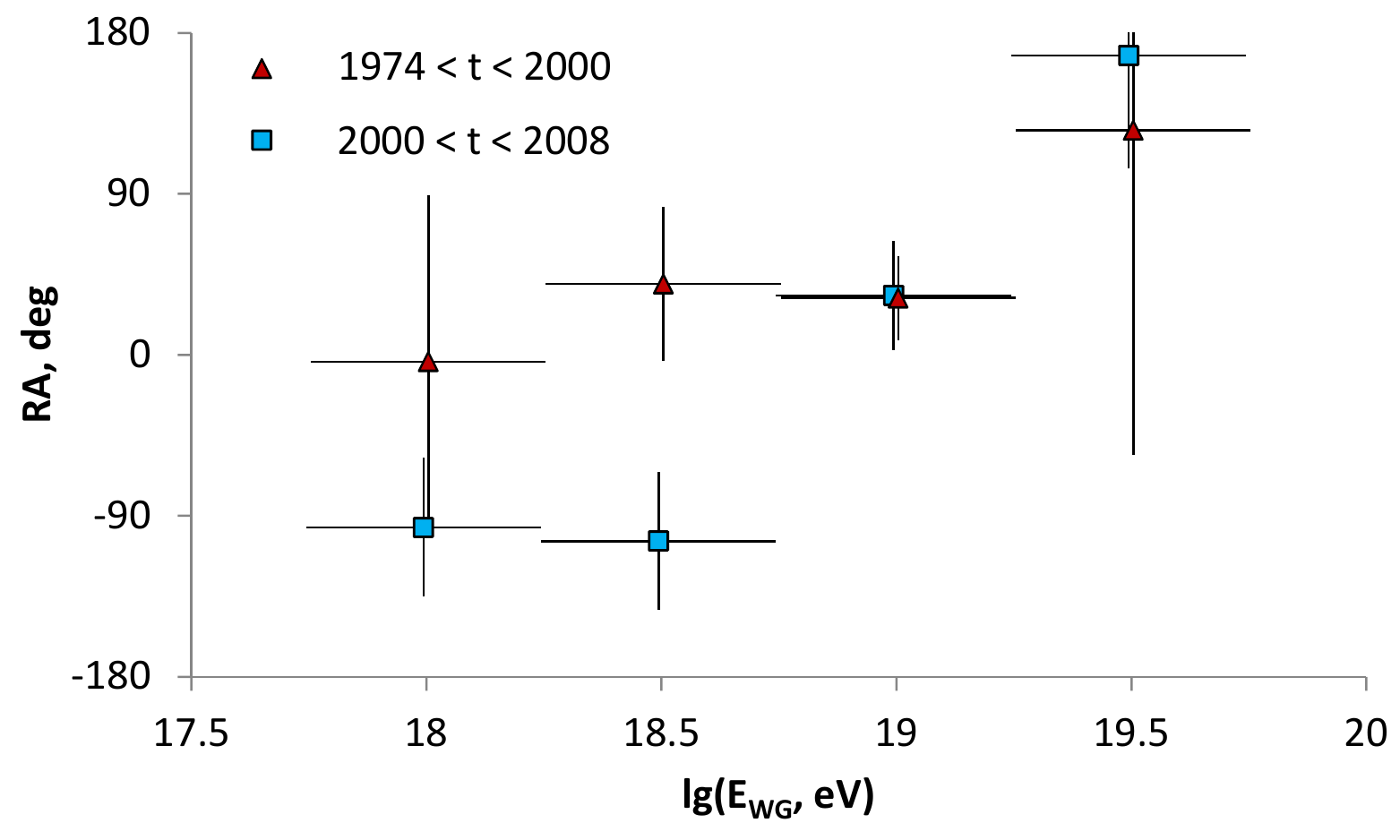}
  \caption{The first harmonic phase as a function of energy. Left panel: comparison of data observed by the Yakutsk array (Ykt), Telescope Array (TA) \cite{TA}, and by the Pierre Auger Observatory (PAO) \cite{PAOphase}. Right panel: comparison of observations at Yakutsk before and after 2000.}
\label{Fig:Phase1}\end{figure}

We compared phases of the first harmonic in energy bins before and after 2000 (Fig.\ \ref{Fig:Phase1}, right panel), and our measurements in these two periods appear to be qualitatively consistent with the PAO conclusion. Although the phase uncertainties in our data are relatively large, the phase of the first harmonic is not randomly distributed, at least above $10$ EeV, but appears to increase gradually in the direction of the anticenter (Fig.\ \ref{Fig:Phase1}, left panel).

Additionally, analysis of the combined datasets of the PAO, TA, and Yakutsk array by the Anisotropy Working Group for the CERN Symposium \cite{WG} also hinted at the same intriguing regularity in the phase of the dipole anisotropy. One conclusion was that larger statistics are needed to investigate these observations.

Our present contribution in strengthening this hint is that the regularity is observed consistently in different time intervals and arrays.

\section{Analysis of variance}
We also tested for the probability of anisotropy in $RA$ distribution using another method, i.e. analysis of variance. This method is able to determine systematic differences between the results of measurements carried out under specific varying conditions \cite{ANOVA}.

The distinctive feature of the isotropic distribution in $RA$, our null hypothesis, $H_0$, is that it has no mean value and the variance $s^2_i=\pi^2/3$ is independent of the trial mean. We define the minimal width of a distribution as $s_m=min\sqrt{\overline{(\alpha-\alpha_0)^2}}$, where the trial mean $\alpha_0$ scans the whole $\alpha\in(0,2\pi)$ interval. The distance between two points can be calculated directly, $|\alpha-\beta|$, or around a circle, $2\pi-|\alpha-\beta|$. We chose the minimal of the two for all pairs. So, under the null hypothesis, the width in degrees is $s_m^i=103.9^0$. On the other hand, if there is a single source, SS, of CRs with width $s\ll s_m^i$ in the isotropic background (our alternative hypothesis, $H_1$), then the aggregate distribution width may be sufficiently narrow depending on the relative luminosity of the source.

For example, if the SS fraction of the total flux is $50\%$, then the resulting width is $s_m=73.5^0$. This can be easily identified by the analysis of variance. Even the flux fraction $10\%$ from the single source resulting in width $s_m=98.6^0$ can be distinguished with a sufficient number of EAS events.

It is convenient to calculate the arrival directions and probabilities of the distribution widths under null and other hypotheses by using the Monte Carlo method. In this case, we can apply the same procedures to form distributions and to calculate variances of the experimental data and random points in the uniform $RA$ distribution. An example of the program is given in Appendix B.

The data of the Yakutsk array were sampled in seven energy bins separated by $\lg E_i=0.25i, i=0,..,6$, where $E_i$ is in units of EeV. For each sample, the minimal width of the $RA$ distribution was found (Table 2) compared with the expected width, $s_m^i$, for an equal number of isotropic events. To avoid the penalty factor in the probability, we divided the dataset into two independent sub-samples: data observed before and after 2000 (the terminal date is 05/31/2000). The former was used to find a bin with minimal distribution width observed; the latter was used to estimate the statistical significance in the energy bin fixed {\it a priori} from the first. The minimal width of $RA$ distribution of 52 EAS events observed in the period 2000--2008 with energies $E\in(10,17.8)$ EeV was $s_m=85^0$. The chance probability that the isotropic distribution width of the 52 events would be less than or equal to the observed width was $P(\leq s_m)=2.2\%$. As a result, we can reject the null hypothesis with at least a $97.8\%$ confidence level.

\begin{figure}[t]\centering
\includegraphics[width=0.5\textwidth]{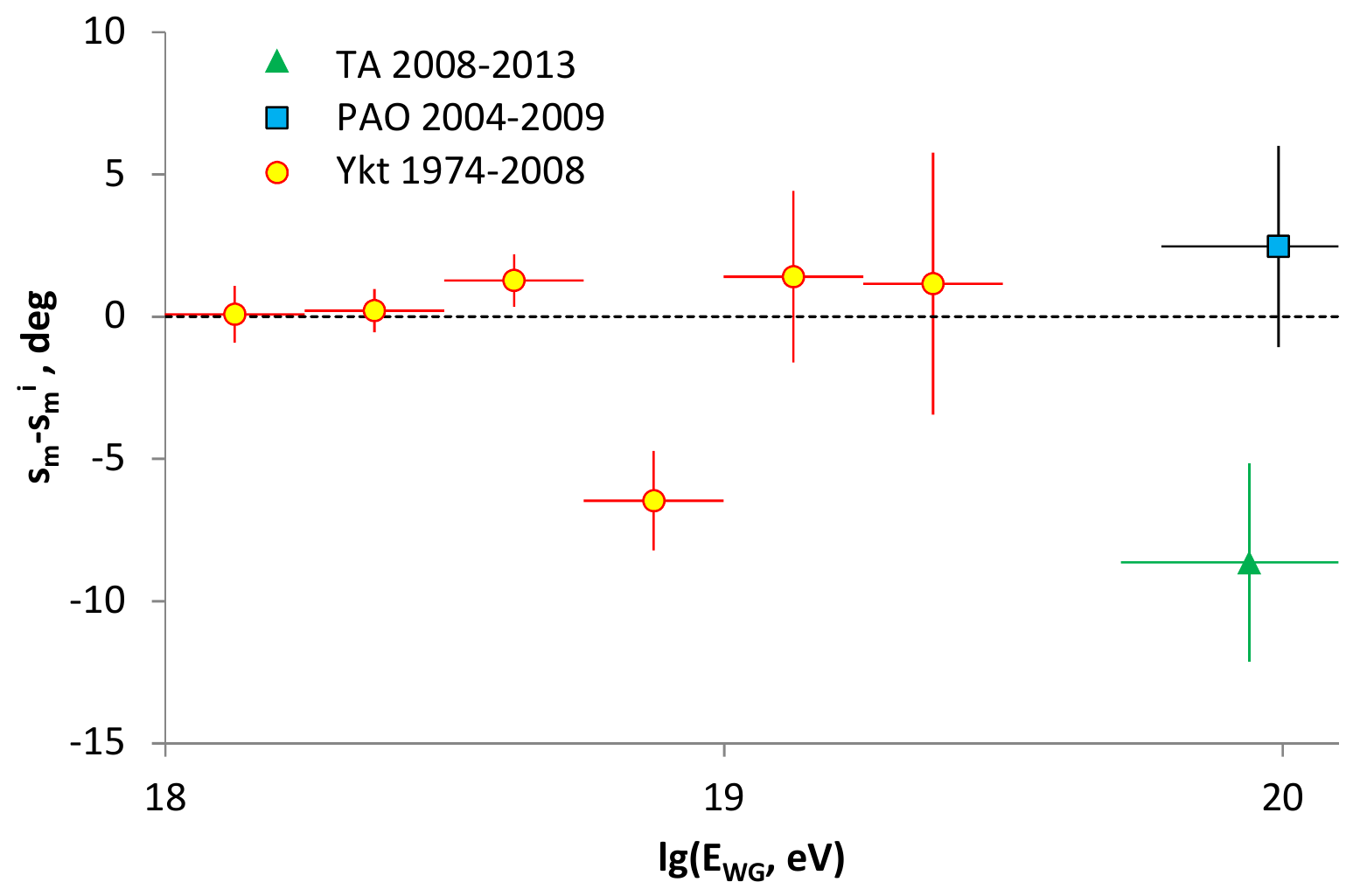}
  \caption{Difference in observed and expected isotropic distribution widths in right ascension. Statistical errors for observed event numbers are shown by the vertical bars.}
\label{Fig:Anova}\end{figure}

In Fig.\ \ref{Fig:Anova}, our results covering the whole period 1974-2008 are shown together with that derived from the published data from PAO \cite{PAOdata} and TA \cite{TA} above $\sim55$ EeV. Consistent with the results of harmonic analysis, there was an energy interval (10, 17.8) EeV where the observed minimal width of the Yakutsk array data was distinctly less than the isotropic expectation.

\begin{table}[t]
\begin{center}
\caption{Minimal width of the right ascension distribution of cosmic rays detected with the Yakutsk array in the two periods. Column heads: event numbers, N; observed, $s_m$, and expected isotropic widths, $s_m^i$; chance probability that isotropic width is less than or equal to observed width, $P$.}
\begin{tabular}{|l|r|r|r|r|r|r|r|r|}\hline
Energy bins, & \multicolumn{4}{c|}{1974-2000}   &\multicolumn{4}{c|}{2000-2008}\\\cline{2-9}
$\lg(E, eV)$ &    N  & $s_m$, & $s_m^i$,&  $P$, &   N   &  $s_m$,&$s_m^i$,& P \\
             &       &   deg  &    deg  &    \% &       &    deg &    deg & \%\\ \hline
 18.00-18.25 & 17063 & 103.66 &  103.43 & 86.06 & 12081 & 103.20 & 103.34 & 28.03 \\ \hline
 18.25-18.50 &  6765 & 103.14 &  103.14 & 45.31 &  3258 & 102.62 & 102.80 & 32.43 \\ \hline
 18.50-18.75 &  2410 & 102.03 &  102.60 & 16.26 &   772 & 100.48 & 101.59 & 14.73 \\ \hline
 18.75-19.00 &   761 & 102.94 &  101.57 & 93.99 &   178 & 102.21 &  99.02 & 96.85 \\ \hline
 19.00-19.25 &   233 &  95.27 &   99.66 &  2.38 &    52 &  84.97 &  94.77 &  2.24 \\ \hline
 19.25-19.50 &    72 &  96.88 &   96.17 & 52.31 &    23 & 100.92 &  89.95 & 99.69 \\ \hline
 19.50-20.00 &    36 &  96.25 &   92.87 & 71.40 &     6 &  82.73 &  74.88 & 66.43 \\ \hline
\end{tabular}
\end{center}
\end{table}

Our alternative hypothesis, $H_1$, has two parameters to fit the experimental data: the source position in $RA$, and the fraction of the total flux produced by the source luminosity. We estimated the most probable $H_1$ parameters fitting the data from Table 2 in the energy range $(10,17.8)$ EeV. The results showed that the SS position is $\alpha\in(15^0,45^0)$, and the ratio of CR flux from SS to the total flux is $I_{SS}/I_{tot}=0.15\pm0.05$.

\section{Conclusions}
We used two methods to examine the $RA$ distribution of CR arrival directions measured with the Yakutsk array: harmonic analysis and analysis of variance. Resultant first and second harmonic amplitudes increased with energy but were consistent with expected amplitudes of the isotropic distribution. In the energy range $(10,31.6)$ EeV, our data observed in the period 1974--2000 exhibited an excess flux with the first harmonic amplitude of $26.4\%$, a chance probability of $0.4\%$, but the second part of the data observed in 2000--2008 had the amplitude $A_1=41.4\%$ with a chance probability of $\sim12\%$. Therefore, we found no significant deviation of the first and second harmonic amplitudes from those in the isotropic distribution, providing a hint of possible anisotropy above $10$ EeV.

Analysis of variance demonstrated the prominent excess in the same energy bin, namely, a significant contraction of the minimal width of the $RA$ distribution of CR arrival directions with respect to the isotropic distribution. Downsizing in width was found in both independent parts of the Yakutsk array data, i.e., data observed before and after 2000. The null hypothesis was rejected at the significance level $\sim98\%$. An alternative hypothesis with a single source in the uniform background flux of CRs was fitted to the observational data in the energy range $(10,17.8)$ EeV with the source position $RA\in(15^0,45^0)$ and the ratio of CR flux from SS to the total flux $I_{SS}/I_{tot}=0.15\pm0.05$.

The first harmonic phase did not appear to be randomly distributed in the interval $(-180^0,180^0)$, as would be expected in the isotropic case, but exhibited a gradual increase with energy in $RA$, at least in the energy interval above $\sim10$ EeV. This behavior is inherent in both Yakutsk observation periods, i.e., before and after 2000, and is in agreement with the possible regularity in the phase of the dipole anisotropy observed by the PAO and TA collaborations.


\section*{Acknowledgments}
We are grateful to the Yakutsk array staff for the data acquisition and analysis. The work is supported in part by the Russian Academy of Sciences (Program 10.2) and RFBR (grants 11-02-00158, 13-02-12036).


\appendix

\section{The first harmonic amplitude of the isotropic right ascension distribution}
Here, we apply the Rayleigh formalism, founded in \cite{Rayleigh}, to illustrate a calculation of the first harmonic amplitude, $A_1$, for $N$ isotropic points in the interval $\alpha\in(0,2\pi)$.

Treating the amplitude as a sum of $N$ vectors of length $|\vec{r_i}|=2/N$ with angle $\alpha_i$, we have the total length, $A_1=|\vec{R_N}|$, which accumulates $N$ equal random steps. The inductive argument about a relationship between $R_{i-1}$ and $R_i$ in a triangle of vectors $\vec{R_i}=\vec{R_{i-1}}+\vec{r_i}$ consists in $\overline{R_i^2}=\overline{R_{i-1}^2}+r_i^2$ due to $\overline{\cos\alpha_i}=0$. Consequently, $\overline{A_1^2}=4/N$.

Asymptotically, the distribution of the amplitude is circular Gaussian according to the central limit theorem: $P(\vec{R_N})=\frac{1}{2\pi\sigma}\exp(-\frac{A_1^2}{2\sigma})$, where $N\gg1$. The mean amplitude is
$$
\overline{A_1}=\frac{1}{\sigma}\int_0^\infty A_1^2\exp(-\frac{A_1^2}{2\sigma})dA_1=\sqrt{2\sigma}\Gamma(1.5)=\sqrt{\frac{\pi\sigma}{2}},
$$
while the mean square of amplitude is
$$
\overline{A_1^2}=\frac{1}{\sigma}\int_0^\infty A_1^3\exp(-\frac{A_1^2}{2\sigma})dA_1=2\sigma\Gamma(2)=2\sigma.
$$
A comparison with the vector length gives the following: $\sigma=2/N$, the mean amplitude $\overline{A_1}=\sqrt{\pi/N}$, and the variance $\overline{A_1^2}-\overline{A_1}^2=(4-\pi)/N$.

The probability of obtaining an amplitude greater than or equal to $A$ is
$$
P(\geq A)=\frac{1}{\sigma}\int_A^\infty A_1\exp(-\frac{A_1^2}{2\sigma})dA_1=exp(-\frac{NA^2}{4}).
$$

\section{A Monte Carlo program to find the minimal width of RA distribution}
The Fortran-90 program below illustrates the minimal width calculation for a distribution of $N$ random points in the right ascension (RA) circle. The function Var computes the minimum variance with the trial mean scanning a circle $(0^0,360^0)$. The mean value and standard deviation of the distribution minimal width are calculated with a sample of size $M$ in the main program W.

Program W;real Q(10000);N=52;M=100000 !  N points sampled M times

 av=0;d=0;do k=1,M

  do i=1,N;Q(i)=RAN(ir)*360.0;enddo;rms=sqrt(Var(Q(1:N),N)) ! N random RA points

 av=av+rms/M;d=d+rms**2/M;enddo;d=sqrt(d-av**2) ! mean \& deviation, degrees

print *,N,M,av,d;end program

function Var(Q,N);integer N;real Q(N) ! Minimal variance of Q(N) points in a circle (0,360)

 smin=1e36;do k=1,36;A=k*10.0 ! A=trial mean

  s=0;do i=1,N;d=abs(Q(i)-A);d=min(d,360.-d);s=s+d**2/N;enddo ! var of Q(N) for A

 if(s$<=$smin)smin=s;enddo;Var=smin;end function.


\end{document}